# Quantitative diffraction imaging using attosecond pulses

G. N. Tran,[1,2,*] Katsumi Midorikawa,[2] and Eiji J. Takahashi[1,3,†]

[1]*Ultrafast Coherent Soft X-ray Photonics Research Team, RIKEN Center for Advanced Photonics, RIKEN, 2-1 Hirosawa, Wako, Saitama 351-0198, Japan*
[2]*Attosecond Science Research Team, RIKEN Center for Advanced Photonics, RIKEN, 2-1 Hirosawa, Wako, Saitama 351-0198, Japan*
[3]*Extreme Laser Science Laboratory, RIKEN Cluster for Pioneering Research, RIKEN, 2-1 Hirosawa, Wako, Saitama 351-0198, Japan*
*\*giang.tran@riken.jp*
*†ejtak@riken.jp*

**Abstract:** We have proposed and developed a method to utilize attosecond pulses in diffraction imaging techniques applied to complex samples. In this study, the effects of the broadband properties of the wavefield owing to attosecond pulses are considered in the reconstruction of images through the decomposition of the broad spectrum into multi-spectral components. This method successfully reconstructs the multi-spectral information of complex samples, probes, and spectral bandwidths using broadband diffraction intensities generated from computational scanning experiments. The results obtained in this research open the opportunities to perform quantitative ultrafast imaging using the attosecond pulses.

## 1. Introduction

The emergence of coherent diffraction imaging (CDI) [1] has resulted from the development of coherent x-ray sources, detectors, and computational phase retrieval algorithms. CDI is a lensless imaging technique that has been used to explore the structure of materials [2] and biological samples [3] using synchrotron radiation (SR) and x-ray free-electron laser (XFEL). In a CDI experiment, a coherent wavefield illuminates a sample, and the diffracted wavefield from the sample is recorded by a detector placed some distance behind the sample. Iterative algorithms [4] are then used, rather than an image forming lens, to reconstruct the image of the sample from the diffraction intensity. Therefore, the quality of the reconstructed image significantly depends on the wavelength of the illuminating wavefield and the numerical aperture of the detector.

The development of table-top sources via high-harmonic generation (HHG) [5-7] facilitates the performance of CDI experiments in the laboratory. In fact, CDI experiments have been successfully performed using monochromatic radiation from table-top sources [8-10], with the obtained results demonstrating the feasibility of imaging samples at high resolution without accessing the large-scale facilities, such as SR and XFEL, to perform the techniques. Recently, the development of attosecond sources [11-17] has highlighted the potential to perform CDI experiments with femtosecond to attosecond temporal resolution. Imaging at high spatial and temporal resolution has brought attention to interesting ultrafast phenomena such as strain dynamics, evolution of structure, phase transitions, or structural variation and interaction mechanisms in cell, which can be explored and applied in material and biological studies [18,19]. Applying attosecond pulses to CDI requires consideration of their characteristics, such as the broad continuum spectra of attosecond sources in HHG and the low photon fluxes involved. Consequently, performing CDI experiments using attosecond pulses is quite challenging.

One possible approach to apply attosecond pulses in CDI experiments is to utilize their broad spectra to increase the photon fluxes incident on the samples. However, applying the broad bandwidth of attosecond pulses violates the requirement of coherence in CDI. As a result, a new method that includes the broadband properties of the beam in the image reconstruction from broadband illumination is necessary. The feasibility of applying broad bandwidth sources in CDI has been demonstrated using HHG, SR and laser sources [20-23] for binary test samples.

Ptychography [24,25], a scanning version of CDI, is a potential candidate for overcoming the limitation of using broad spectrum sources. In ptychography, the sample is scanned across the illuminating beam with an overlapping between two adjacent positions, as illustrated in Fig. 1. Sets of diffraction intensities are recorded from different positions of the sample, producing a data redundancy in the reconstruction that is not present in the conventional CDI. The image reconstruction is then performed iteratively from one position to another. Consequently, ptychography is a robust and promising method for overcoming the negative effects caused by broadband illumination in CDI. This method has been applied in the reconstruction of test patterns in simulation and experiments from broadband illumination [26-29].

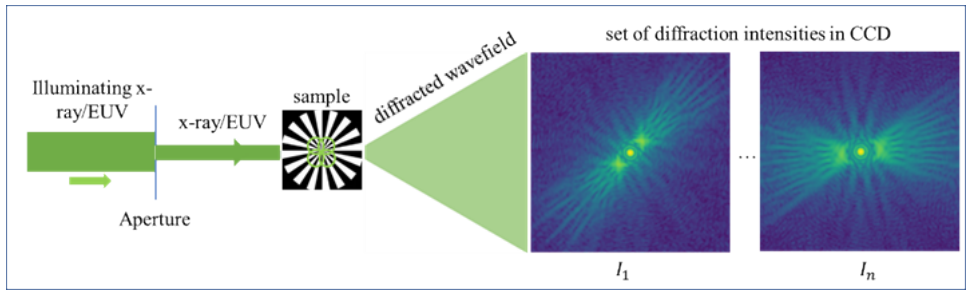

Fig. 1. Illustration of a ptychographic experiment.

Using the ptychographic approach, we have proposed a method to apply the broad spectrum of attosecond pulses in quantitative diffraction imaging. As opposed to the method described in [27], this method applies a scaling rule [20] based on the broad spectrum of the attosecond pulses to incorporate the effects of broadband illumination in the reconstruction and quantitative reconstructs the spectral information of complex samples at different energies. This approach is an alternative and simpler than the method described in [27] while spectrally quantitative information of complex sample can still be reliably recovered. Validation is numerically performed using complex samples that undergo absorption and phase changes at the illuminating energies. The obtained results demonstrate that the broad spectrum of the attosecond pulses generated from HHG processes can be applied to obtain high-quality images and spectral information for complex samples without the needs to repeat measurements using monochromatic illumination. Meanwhile, the spectral bandwidth of the illuminating source is also retrieved in the image reconstruction. The obtained results imply that the method can be applied to other sources that have broad spectra such as SR, XFEL, and x-ray laser plasma sources,…

## 2. Method

### 2.1 Simulated broad spectrum

The broad spectrum used for the simulation was generated from the experimental results [15,16]. This spectrum was chosen to perform a simulation consistent with the physical conditions of the experimental source. In addition, recent experiments have demonstrated an isolated attosecond pulse (IAP) with a high pulse energy of 0.24 $\mu J$ and pulse duration of 226 as [16]. These results suggest the feasibility of single-shot imaging with high temporal resolution using an IAP. Therefore, this research aims to investigate the viability of applying the IAP to diffraction imaging. The broad spectrum has a continuous distribution of wavelengths ranging from 17.5 to 24.6 nm, which corresponds to energies of 50.5 to 71 eV with a central peak of 61 eV. This energy spectrum is equivalent to the measured spectrum of an IAP, as shown in Fig. 2 of [16].

The broad spectrum is then decomposed into 12 various monochromatic photon energies (wavelengths) with their own spectral intensities and no bandwidth, as shown in Fig. 2 and Table S1 of the supplemental document, to model the effects of the broad bandwidth on the diffraction intensities. To the best of our knowledge, there are currently no criteria regarding how fine the spectral decomposition should be to fully describe the properties of the broad spectrum. Logically, a finer sampling would achieve a more optimal approximation to the actual spectrum; yet, more spectral components from a finer sampling may cause difficulties in the image reconstruction process. A more thorough investigation of this issue is beyond the scope of this research and will be pursued in a future study. In this research, a finer decomposition is also performed, nevertheless, there is insignificant difference for the reconstructed results. Therefore, the decomposition of the broad spectrum into 12 monochromatic photon energies is chosen for this research.

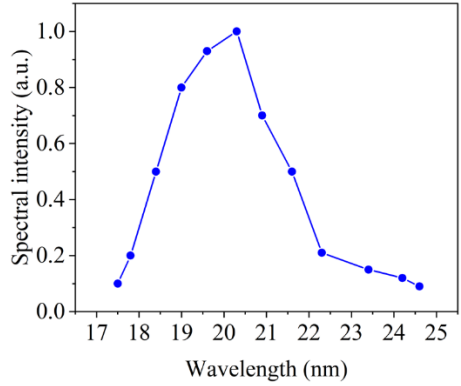

Fig. 2. Simulated broad spectrum.

### 2.2 Simulated complex sample

The samples used in this research are complex samples containing amplitudes and phases representing the absorption and phase components of the refractive indices of the samples. The 12 monochromatic photon energies of the decomposed broad spectrum correspond to the 12 independent spectral amplitudes and phases of the complex sample. At each wavelength, the complex sample is defined as [30],

$$O(\vec{r},\lambda) = A(\vec{r},\lambda) \times exp[i\phi(\vec{r},\lambda)], \qquad (1)$$

where $O(\vec{r},\lambda)$, $A(\vec{r},\lambda)$ and $\phi(\vec{r},\lambda)$ are the transmission function, amplitude, and phase of the complex sample, respectively. $\vec{r}$ and $\lambda$ are the two-dimensional coordinates at the sample plane and the wavelength of the illuminating wavefield, respectively.

The amplitude and phase are, respectively, calculated as,

$$A(\vec{r},\lambda) = exp[-k(\lambda) \times \beta(\lambda) \times t(\vec{r})] \qquad (2)$$

$$\phi(\vec{r},\lambda) = k(\lambda) \times \delta(\lambda) \times t(\vec{r}), \qquad (3)$$

where $t(\vec{r})$ represents the thickness map of the sample and $k(\lambda) = 2\pi/\lambda$. $\delta(\lambda)$ and $\beta(\lambda)$ are respectively calculated as [31],

$$\delta(\lambda) = (n_a \times r_e \times \lambda^2 \times f_1)/2\pi \qquad (4)$$

$$\beta(\lambda) = (n_a \times r_e \times \lambda^2 \times f_2)/2\pi, \qquad (5)$$

where $n_a$ is the atomic density of the sample material, $r_e$ is the Bohr radius of an atom, $f_1$ and $f_2$ are the first and second atomic form factors of atom, respectively.

In the first simulation, a resolution test pattern, shown in Fig. 3a, was chosen as the simulated sample with a thickness of 100 nm. The thickness map of the simulated sample is then generated by multiplying the sample by the thickness. Finally, the spectral thickness maps are generated by scaling the thickness map with the wavelengths in the decomposed broad spectrum, as described in [20], nevertheless, the scaling is performed in the real space to retain the quantitative information of complex samples in this work. The generated thickness map from the central component ($\lambda$ = 20.3 nm) which has the highest intensity is shown in Fig. 3b. The other spectral thickness maps are included in Fig. S1 of the supplemental document.

Nickel was chosen as the material of the sample. The values of $f_1$ and $f_2$ are obtained from [31], yielding $\delta(\lambda)$ and $\beta(\lambda)$ through Eqs. (4) and (5). The spectral amplitudes and phases of the complex sample for each wavelength are calculated using Eqs. (2) and (3). Figures 3c and 3d show the amplitude and phase of the complex sample, respectively, for the central spectral component ($\lambda$ = 20.3 nm). The other generated spectral amplitudes and phases for this sample are respectively shown in Figs. S2 and S3 of the supplemental document.

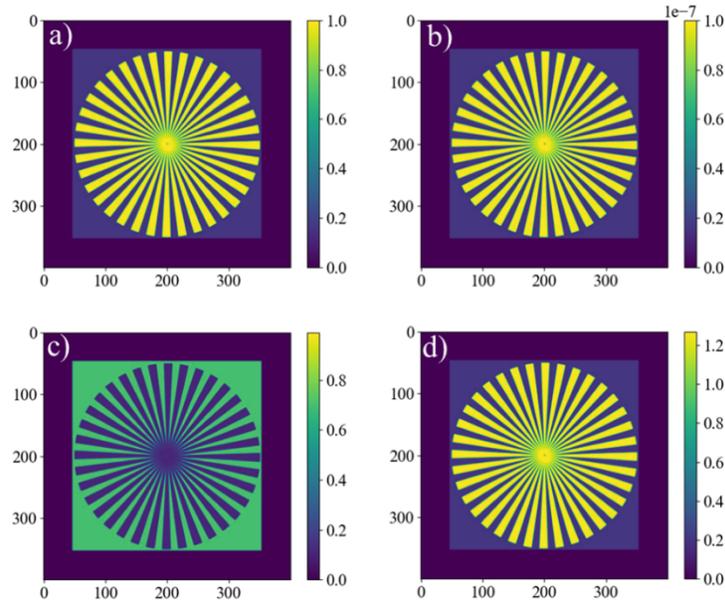

Fig. 3. Simulated complex sample. (a) Resolution test pattern used for the sample, (b) thickness map of the sample obtained by multiplying (a) by the thickness at $\lambda$ = 20.3 nm, (c) and (d) amplitude and phase at $\lambda$ = 20.3 nm.

### *2.3 Simulated probe*

In the experimental measurements, an aperture is usually used to define the probe coming the sample. Therefore, for greater fidelity, a simulated probe is generated by propagating a wavefield from a binary circular aperture, as shown in Fig. 4a, to the sample plane. Similarly, the probe is scaled with spectral wavelengths to generate spectral probes. Consequently, 12 spectral probes were obtained from the 12 spectral wavelengths. The the sample is placed at 10 $\mu$m from the aperture and the propagation is performed using Fresnel propagation [32]. Figures 4b and 4c show the amplitude and phase, respectively, of the central probe at a wavelength of 20.3 nm. The other spectral probes are shown in Figs. S4 and S5 of the supplemental document.

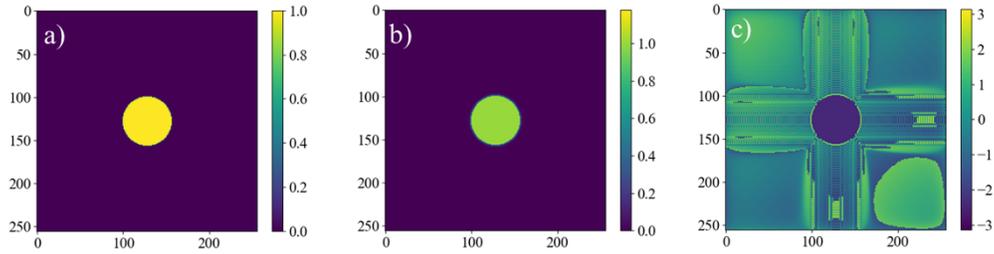

Fig. 4. Probe used in the simulation. (a) Binary circular aperture, (b) and (c) amplitude and phase of the central probe ($\lambda$ = 20.3 nm) at the sample plane.

## 2.4 Generation of the broadband diffraction intensity

The broadband diffraction intensity from each scanning position is generated as follows,

Step 1: Generate the spectral exit surface waves (ESWs) by multiplying the spectral probes and spectral transmission functions,

$$\Psi_j(\vec{r},\lambda_l) = P(\vec{r} - \vec{R}_{s(j)},\lambda_l) \times O_j(\vec{r},\lambda_l) \tag{6}$$

where $j$ and $l$ are the scanning position and spectral component, respectively. $P(\vec{r} - \vec{R}_{s(j)},\lambda_l)$ represents the spectral probe, and $\vec{R}_{s(j)}$ is the relative shift between the probe and sample [33].

Step 2: Propagate the spectral ESWs to the far-field,

$$\hat{\Psi}_j(\vec{q},\lambda_l) = \Im^{+1}[\Psi_j(\vec{r},\lambda_l)], \tag{7}$$

where $\vec{q}$ and $\Im^{+1}$ represent the reciprocal space coordinates and forward Fourier transform, respectively.

Step 3: Compute the broadband diffraction intensity for each scanning position,

$$I_j(\vec{q}) = \sum_{l=0}^{n-1} \eta_l \times \left|\hat{\Psi}_j(\vec{q},\lambda_l)\right|^2, \tag{8}$$

where $\eta_l$ is the spectral intensity of the spectral component $l$ in the spectrum and $n$ is the number of spectral components in the decomposed broad spectrum.

The diffraction intensities from the other positions of the sample are obtained by scanning the sample across the beam and repeating steps 1-3.

To provide greater validity, noise is added to each pixel of the diffraction intensities. The noise model is chosen to be the Poisson distribution, which is proportional to the photon number [34]. The chosen photon numbers are based on the pulse energy of the IAP which has approximately $10^{10}$ photons/shot [16]. Owing to the sample absorption and quantum efficiency of the CCD camera, the photon number recorded by the CCD camera should be less than $10^{10}$ photons. Thus, the photon number of $10^6$, $10^8$ and $10^{10}$ photons are respectively chosen in this study to represent different levels of noise. Higher photons imply better signal to noise ratios and vice versa. Figure 5 shows two diffraction intensities from two scanning positions of the complex sample before and after adding noise. In this research, different noise levels are added to the broadband diffraction intensities and some of diffraction intensities which noise added are shown in Fig. S6 of the supplemental document. The overlapping between two consecutive scanning positions, which plays an important role in the reconstruction is chosen to be 96%. Compared to monochromatic ptychography, this overlapping ratio is higher to account for the effects of broadband illumination. The multi-spectral information requires higher overlapping to be obtained from broadband diffraction intensities.

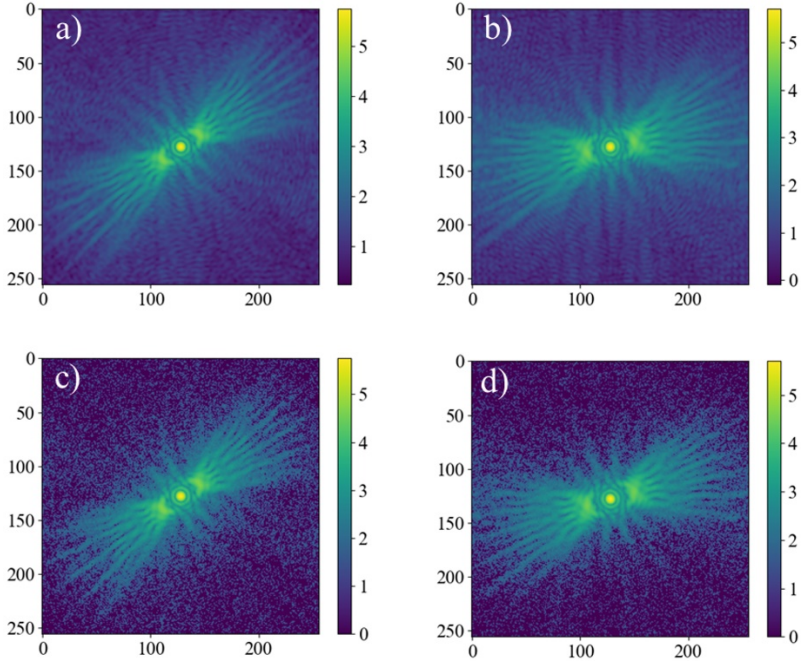

Fig. 5. Broadband diffraction intensities at two scanning positions. (a) and (b) before adding noise, (c) and (d) corresponding intensities after adding noise. The noise level is proportional to $10^6$ photons. The intensities are given in logarithmic scale.

*2.5 Image reconstruction process*

The image reconstruction for the ptychographic method using broadband illumination is the broadband extension of the method given in [33] and is described as follows,

Step 1: The initial guesses of samples $O_j^i(\vec{r},\lambda_l)$ and probes $P_j^i(\vec{r}-\vec{R}_{s(j)},\lambda_l)$ are generated. In this broadband illumination, the probes are then scaled with spectral wavelengths to account for the broadband effects, where $i$ is the iteration number.

Step 2: Multiply the spectral probes by the samples to generate the spectral ESWs,

$$\Psi_j^i(\vec{r},\lambda_l) = P_j^i(\vec{r}-\vec{R}_{s(j)},\lambda_l) \times O_j^i(\vec{r},\lambda_l), \qquad (9)$$

Step 3: Propagate the spectral ESWs to the far-field plane using the forward Fourier transform,

$$\hat{\Psi}_j^i(\vec{q},\lambda_l) = \mathfrak{I}^{+1}[\Psi_j^i(\vec{r},\lambda_l)]. \qquad (10)$$

Step 4: Calculate the estimated broadband diffraction intensity,

$$I_j^{ei}(\vec{q}) = \sum_{l=0}^{n-1} \eta_l \times \left|\hat{\Psi}_j^i(\vec{q},\lambda_l)\right|^2. \qquad (11)$$

Step 5: Apply the modulus constraint to the spectral wavefields,

$$\hat{\Psi}_j^{i\prime}(\vec{q},\lambda_l) = \hat{\Psi}_j^i(\vec{q},\lambda_l) \times \frac{I_j^m(\vec{q})}{I_j^{ei}(\vec{q})}, \qquad (12)$$

where $I_j^m(\vec{q})$ represents the measured diffraction intensities and in this simulation, they are generated in Sec. 2.4.

Step 6: Propagate the spectral wavefield in Step 5 back to the sample plane using the backward Fourier transform,

$$\Psi_j^{i'}(\vec{r},\lambda_l) = \mathfrak{F}^{-1}\left[\hat{\Psi}_j^{i'}(\vec{q},\lambda_l)\right], \tag{13}$$

where $\mathfrak{F}^{-1}$ is the backward Fourier transform.

Step 7: Update the spectral probes and samples,

$$P_{j+1}^i(\vec{r},\lambda_l) = P_j^i(\vec{r},\lambda_l) + \gamma \frac{O_j^{i*}(\vec{r}+\vec{R}_{s(j)},\lambda_l)}{\left|O_j^i(\vec{r}+\vec{R}_{s(j)},\lambda_l)\right|^2_{max}}\left(\Psi_j^{i'}(\vec{r},\lambda_l) - \Psi_j^i(\vec{r},\lambda_l)\right) \tag{14}$$

$$O_{j+1}^i(\vec{r},\lambda_l) = O_j^i(\vec{r},\lambda_l) + \alpha \frac{P_j^{i*}(\vec{r}-\vec{R}_{s(j)},\lambda_l)}{\left|P_j^i(\vec{r}-\vec{R}_{s(j)},\lambda_l)\right|^2_{max}}\left(\Psi_j^{i'}(\vec{r},\lambda_l) - \Psi_j^i(\vec{r},\lambda_l)\right). \tag{15}$$

Steps 1-7 are performed for all spectral components, and the reconstruction is repeated for the next scanning position, using the updated spectral probes and samples from the previous scanning position as initial guesses.

The spectral intensities were updated along with the reconstruction of the spectral probes and samples as [27],

$$\eta(\lambda_l) = \Sigma_{\vec{r}}\left|P(\vec{r}-\vec{R}_{s(j)},\lambda_l) \times O(\vec{r},\lambda_l)\right|^2. \tag{16}$$

Contrary to the method described in [33], the image reconstruction method described here is performed using multi-spectral wavefields from the broad spectrum. Furthermore, to account for the variation at different wavelengths, the initial guesses of spectral probes were scaled with the spectral wavelengths in the sample plane. Using this approach, the spectral information of complex samples can be retained and reconstructed at different spectral wavelengths.

## 3. Results and discussion

The image reconstruction procedure was applied to reconstruct the probes, sample images, spectral bandwidths, and sample spectral information from a set of diffraction intensities. At the beginning of the reconstruction, two-dimensional arrays of one are used for the initial guesses of the sample. In realistic experiments, the initial guesses of the probes can be modeled by propagating a wavefield from the aperture to the sample planes. Therefore, a binary circular aperture is generated and scaled with spectral wavelengths, followed by a propagation to the sample plane to initialize the estimate of the spectral probes. Figures 6 and 7 show several reconstructed spectral amplitudes and phases of the sample and probe, respectively, after completion of the reconstruction for a photon number of $10^8$ photons and overlapping of 96%. The other reconstructed spectral amplitudes and phases of the sample and probe are shown in Figs. S7, S8, S9, and S10 of the supplemental document. The results indicate that the method can reconstruct multi-spectral amplitudes and phases from one experiment. A comparison of the results from different multi-spectral components demonstrates that the quality of the reconstructed spectral amplitudes and phases is degraded when using diminished spectral intensities. In particular, the result at the central wavelength, which has the highest spectral intensity, is better than the results for the other spectral components. The image quality deteriorates with the distance from the central component.

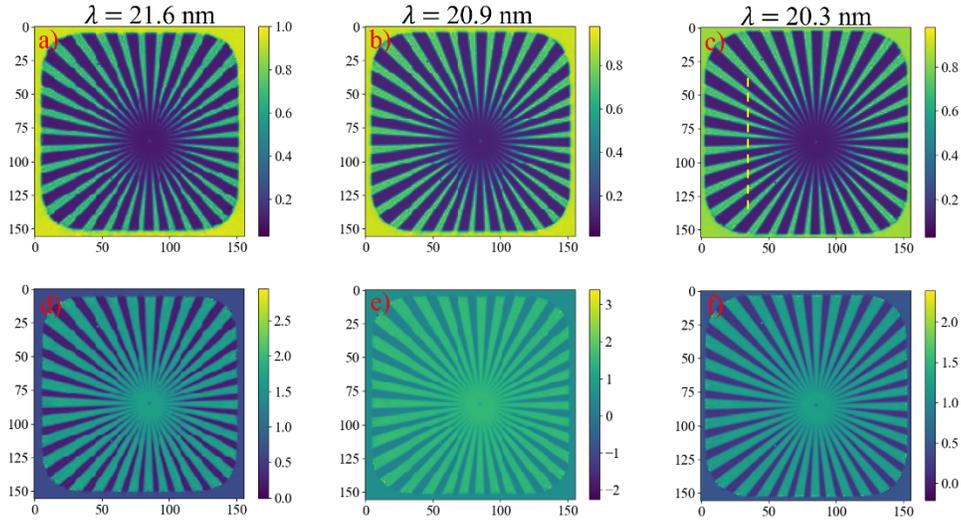

Fig. 6. Reconstructed results at different spectral wavelengths. (a), (b) and (c) reconstructed amplitudes; (d), (e) and (f) corresponding reconstructed phases of the sample. The yellow dashed line shown in (c) indicates the line profile used for quantitative evaluation. The photon number of $10^8$ and the overlapping of 96% are used.

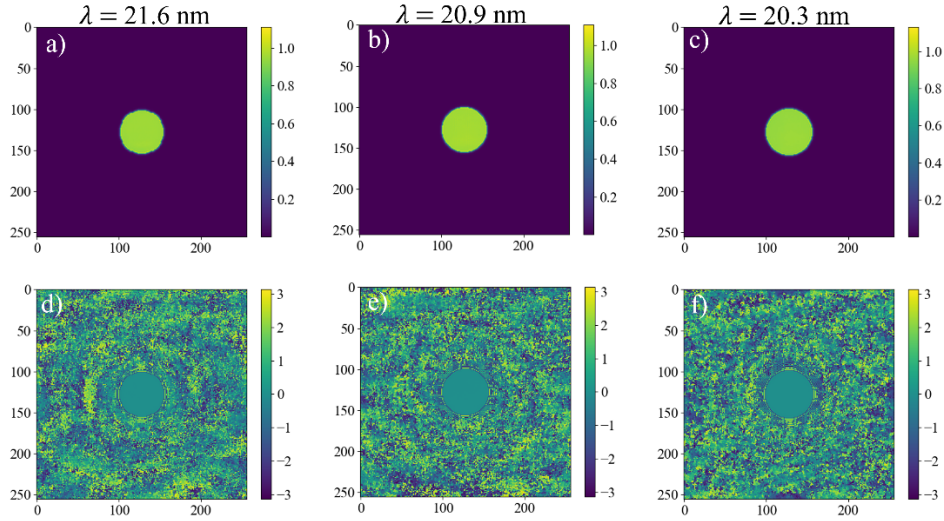

Fig. 7. Reconstructed probes at different spectral wavelengths obtained along with the results shown in Fig. 6. (a), (b) and (c) reconstructed amplitudes; (d), (e) and (f) corresponding reconstructed phases.

Furthermore, the simulation was performed with $10^6$ and $10^{10}$ photons in the illuminating wavefield and it is realized that when the photon number increases, more high-quality spectral images are obtained, and their quality is also improved. In addition, the obtained results show that the structure of the sample can be reliably retrieved, even when the photon number is low, as indicated by the example of $10^6$ photons shown in Fig. S11 of the supplemental document. These results promote the opportunity for performing ultrafast ptychography of vibrating samples [35] using broadband femtosecond or attosecond sources with low photon numbers. In fact, the results obtained with low photon fluxes in this work imply that if only the structure of the sample is desired, the diffraction data can be measured from a short exposure time using ultrafast sources, yielding the sample's structure from the reconstructed result of the central wavelength, while neglecting the information from the other spectral components. In this study, the other spectral reconstructed images for $10^6$ and $10^{10}$ photons show a similar variation to the

$10^8$ photons image; in particular, high-quality images are obtained at high intensities. However, the degradation of the image quality at low spectral intensities for $10^6$ photons occurs faster than that for $10^8$ and $10^{10}$ photons. Comparing the results of $10^8$ and $10^{10}$ photons shows that the changes in the reconstructed results are not significant. This demonstrates that this method can be applied to low photon sources, such as table-top femtosecond and attosecond sources [8-10, 11-16].

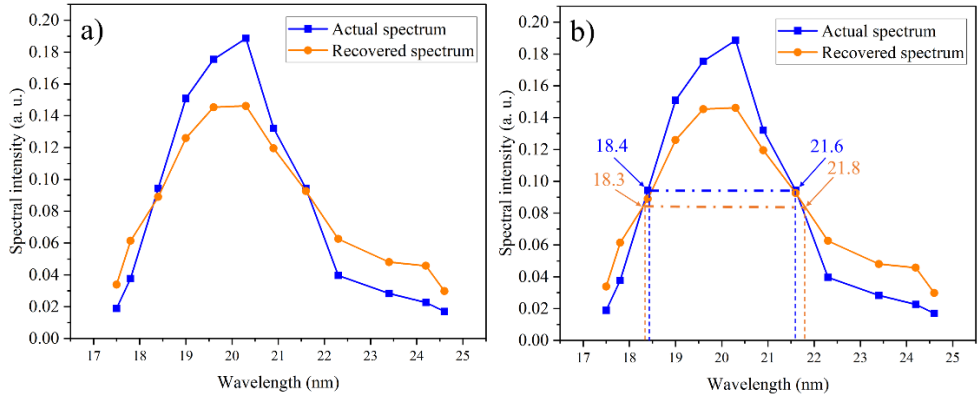

Fig. 8. Comparison of the actual and recovered broad spectra obtained along with the results shown in Fig. 7. (a) Comparison of the actual and the recovered spectra, (b) estimation of the actual and recovered bandwidths: actual bandwidth = (21.6-18.4) = 3.2 nm, recovered bandwidth = (21.8-18.3) = 3.5 nm which is approximately 9.4% difference from the actual bandwidth.

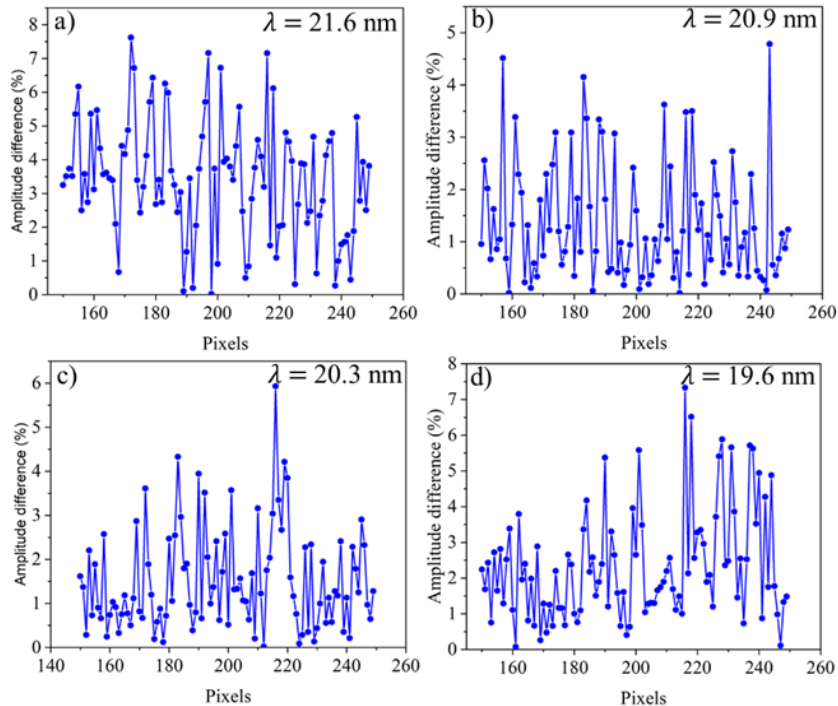

Fig. 9. Quantitative evaluation of reconstructed spectral amplitudes of the sample. (a), (b), (c) and (d) the difference of the reconstructed spectral amplitudes from the actual values. The data is taken across the dashed line shown in Fig.6c.

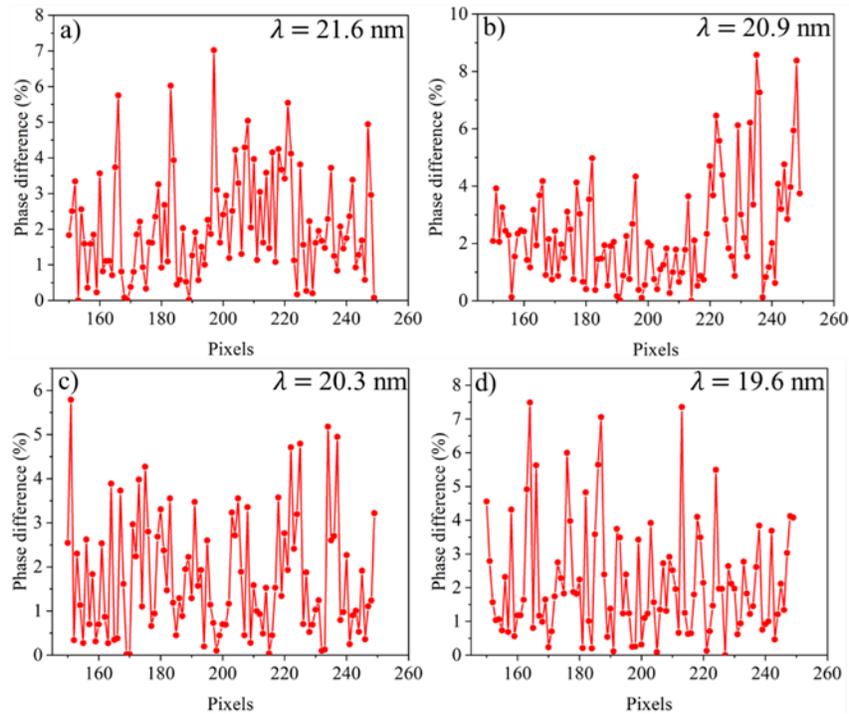

Fig. 10. Quantitative evaluation of reconstructed spectral phases of the sample. (a), (b), (c) and (d) the differences of the reconstructed spectral phases from the actual values. The data is taken from the same line as Fig. 9.

Another quantity that can be obtained from this method is the bandwidth of the broad spectrum. In diffraction imaging, the quality of the reconstructed images is influenced by the spectral bandwidth of the source [36,37]. Consequently, the recovered spectral intensities are then used to estimate the spectral bandwidth. Figure 8 shows a comparison of the recovered and actual spectra obtained from the spectral intensities. It is evident that there are differences in the spectral intensities of the two spectra. However, the beam shape is consistent, and, more importantly, the recovered bandwidth is in good agreement with the actual values. In particular, the bandwidths of the actual and recovered spectra are estimated by their full width at half maximum, yielding bandwidths of the actual and recovered spectra of approximately 3.2 and 3.5 nm, respectively, representing an approximately 9.4% difference between the two values. Consequently, the agreement between the two bandwidths implies that the broadband diffraction intensities generated from the spectrum with the recovered bandwidth coincide with the actual intensities.

In addition, one of the advantages of ptychographic experiments using broadband illumination is the possibility of recovering spectral information of the sample. It has been shown that the spectral absorbances and phases of the sample can be obtained by repeating ptychographic experiments at different energies [38,39]. Such spectral information can be combined with the information obtained from other techniques, such as X-ray Absorption Fine Structure (XAFS) to recover the physical and chemical properties of the sample [38,39]. The results obtained from broadband illumination in this study demonstrate that the multi-spectral information of the sample can be reliably recovered without repeating multiple experiments at different energies. The recovered results are compared with the actual values by taking a line profile across the image, as shown in Fig. 6c. The recovered and actual values are shown to be in good agreement for four different spectral amplitudes and phases of the sample using one ptychographic experiment. In particular, their differences are within approximately 10% for the reconstructed amplitudes, shown in Fig. 9, and phases, shown in Fig. 10. As a result, the absorbance information [38] of the sample at different energies can be obtained with a single ptychographic measurement, rather than having to repeat the process at different times for a single energy. Figure 11 shows the agreement between the actual and recovered absorbances from the four different spectral amplitudes shown in Fig. 9. The absorbances are calculated from one nonzero point across the dashed line shown in Fig. 6c. This result implies that the exposure time and radiation dose of the experiment can be significantly reduced using broadband illumination, which is beneficial for experiments with dose-sensitive samples such as biological samples [40].

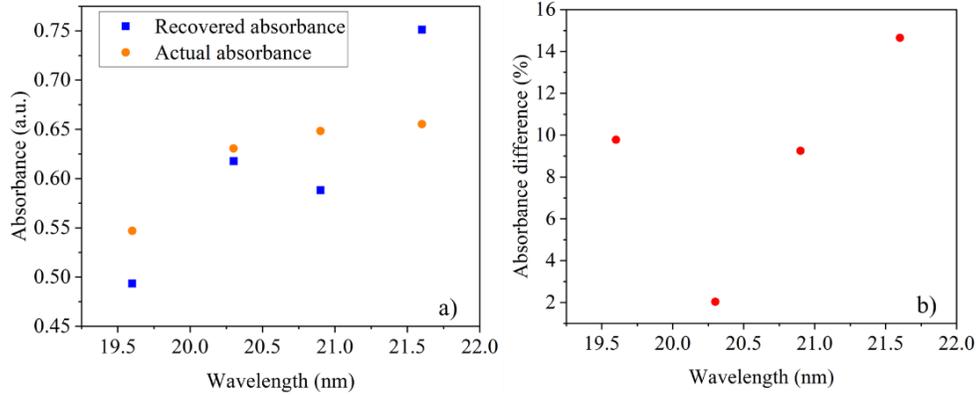

Fig. 11. Recovery of spectral absorbances from reconstructed spectral amplitudes of the sample. (a) Comparison of the recovered and actual spectral absorbances at different spectral wavelengths, and (b) the corresponding differences between the recovered and actual results shown in (a).

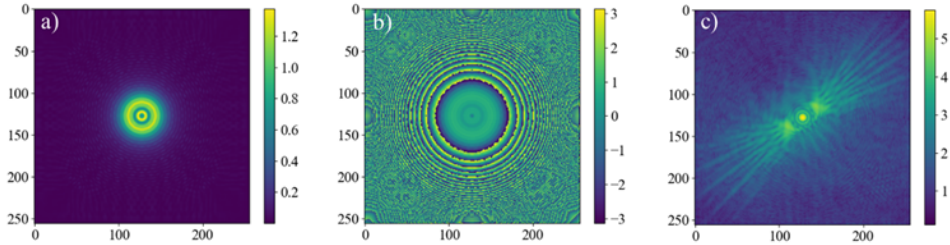

Fig. 12. Simulation with another type of probe. (a) and (b) amplitude and phase of the central probe, (c) broadband diffraction intensity (in logarithmic scale) with the photon number of $10^8$ photons at a scanning position using the sample shown in Fig. 3.

The simulation is also performed using another type of profile, accounting for a different illuminating probe in realistic conditions. Figures 12a and 12b show the amplitude and phase, respectively, of the central spectral probe used in this simulation. The other spectral probe amplitudes and phases are shown in Figs. S12 and S13 of the supplemental document, respectively. The probes were generated by increasing the propagation distance between the circular aperture and sample planes to approximately 1 mm. The broadband diffraction intensities were then generated using the procedure described in Sec. 2.4. Figure 12c shows the broadband diffraction intensity generated from a scanning position on the complex sample shown in Fig. 3, using this type of probe and a photon number of $10^8$ photons. One of the reconstructed spectral amplitudes and phases are shown in Fig. 13. The other results are shown in Figs. S14 and S15 of the supplemental document. In this simulation, only successfully reconstructed spectral amplitudes and phases are considered and shown. The obtained results show that it is more difficult to quantitatively reconstruct the spectral amplitudes and phases of the sample for this type of probe. In particular, when investigating the difference in the reconstructed spectral amplitudes and phases from the actual values, the number of spectral amplitudes and phases that are in good agreement with the original ones is reduced, 3 vs. 4 in the previous simulation. The successfully reconstructed spectral amplitudes and phases of the probe obtained along with the results shown in Fig. 13 are shown in Fig. S16 of the supplemental document. The reconstructed spectral samples and probes demonstrate that reliable results are obtained only at high spectral intensities, as described in the previous simulation. Meanwhile, when investigating the reconstructed broad spectrum, shown in Fig. S17 of the supplemental document, some sudden changes are demonstrated in the recovered spectrum, especially at low intensities. However, these unexpected changes occur at low intensities and do not significantly affect the reconstructed results. In addition, the recovered bandwidth shows good agreement with the actual bandwidth. Therefore, it can be concluded that even though it is more difficult to use this type of probe to reconstruct spectral information in broadband ptychography, reliable results can still be obtained with limited spectral information, compared to those obtained from a well-defined beam. Consequently, in experimental measurements, more localized and well-defined probes are necessary to optimize the reconstructed results, which can be obtained by placing the sample near the circular aperture.

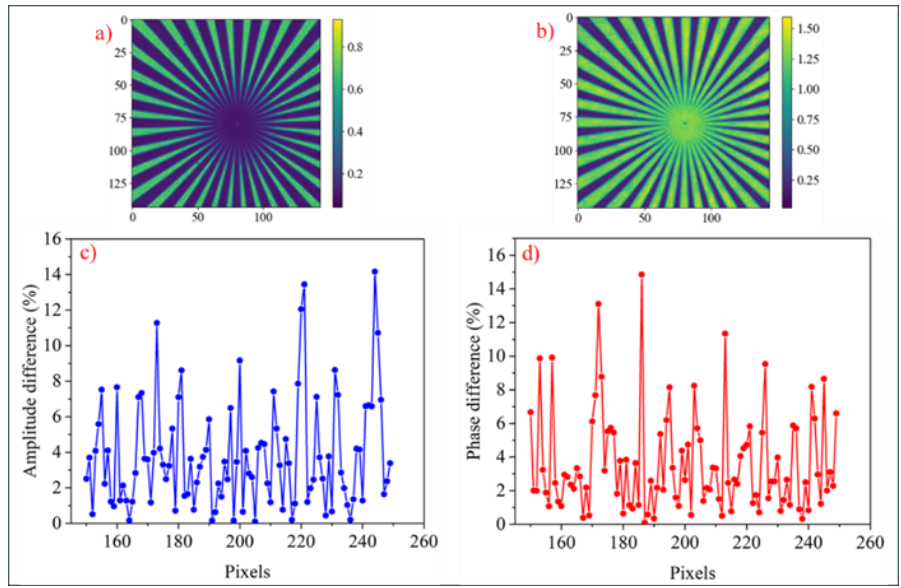

Fig. 13. Results obtained from simulation with the probe shown in Fig. 12. (a) and (b) reconstructed amplitude and phase of the sample for the central wavelength of 20.3 nm, (c) and (d) their difference from the actual values. The data is taken from the line profile shown in Fig. 6c.

Another type of sample demonstrating continuous variations of amplitude and phase is also applied through the simulation. In this simulation, a digital standard test image is used for the simulated sample. Spectral amplitudes, phases of the sample and broadband diffraction intensities are generated using the procedures given in Secs. 2.2 and 2.4. The spectral probes shown in Fig. 4 are used for this investigation. Figure 14 shows the central amplitude and phase of the simulated sample and the broadband diffraction intensity at a scanning position with a photon number of $10^8$ photons. As the procedure used to perform this simulation is the same as that used in the previous simulation, the amplitudes and phases of the other spectral probe and sample are chosen not to be shown.

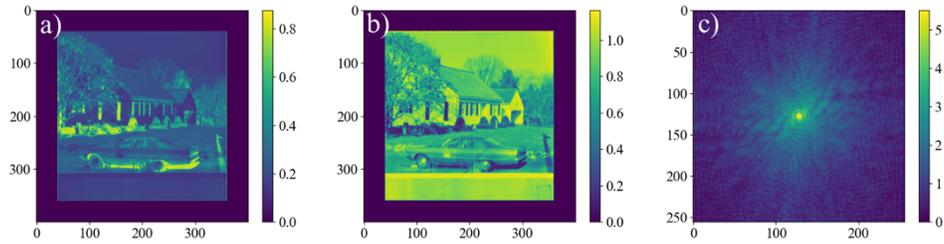

Fig. 14. Simulation using another type of sample. (a) and (b) sample's amplitude and phase generated from the central wavelength (20.3 nm), (c) broadband diffraction intensity (in logarithmic scale) with the photon number of $10^8$ at a scanning position.

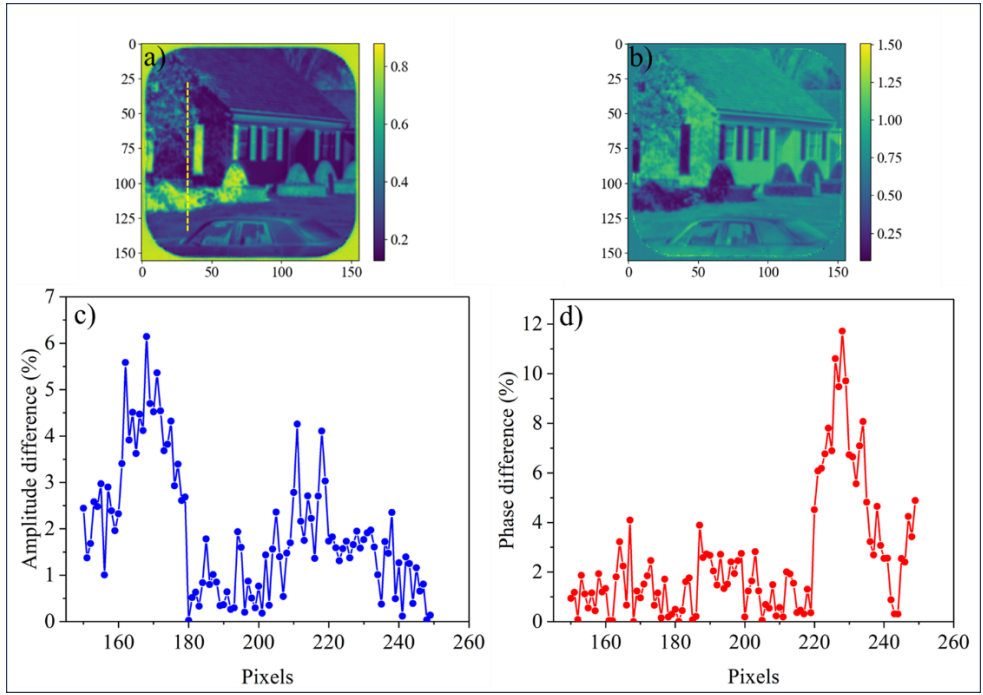

Fig. 15. Results obtained from the simulation shown in Fig. 14. (a) and (b) reconstructed amplitude and phase of the sample for the wavelength of 20.3 nm, (c) and (d) their difference from the actual values. The data is taken across the yellow dashed line shown in (a).

The reconstructed spectral amplitude and phase of the sample for the central wavelength are shown in Fig. 15. Figure 16 shows a quantitative comparison of the other spectral amplitudes and phases of the sample with the actual values. The other results are shown in Figs. S18 and S19 of the supplemental document. These results demonstrate that the conclusion obtained for the previous simulation also applies for this simulation. The number of high-quality reconstructed spectral amplitudes and phases of the sample is four, which is unchanged from the results of the resolution test sample. In particular, there is good agreement between the reconstructed and actual amplitudes and phases of the sample for different wavelengths. A quantitative comparison of the reconstructed spectral amplitudes and phases of the sample with the actual values shown in Figs. 15 and 16 indicates the reliability of the reconstructed results. Therefore, the obtained results imply that the method is reliable for reconstructing the spectral amplitudes and phases for this type of sample. The reconstructed spectral probes are also in good agreement with the actual probes for high spectral intensities, shown in Figs. S20 and S21 of the supplemental document. Furthermore, the recovered spectral bandwidth, shown in Fig. S22 of the supplemental document, coincides with the actual value. Thus, it can be concluded that the method is also reliable for continuous-valued samples in practical applications such as material and biological samples.

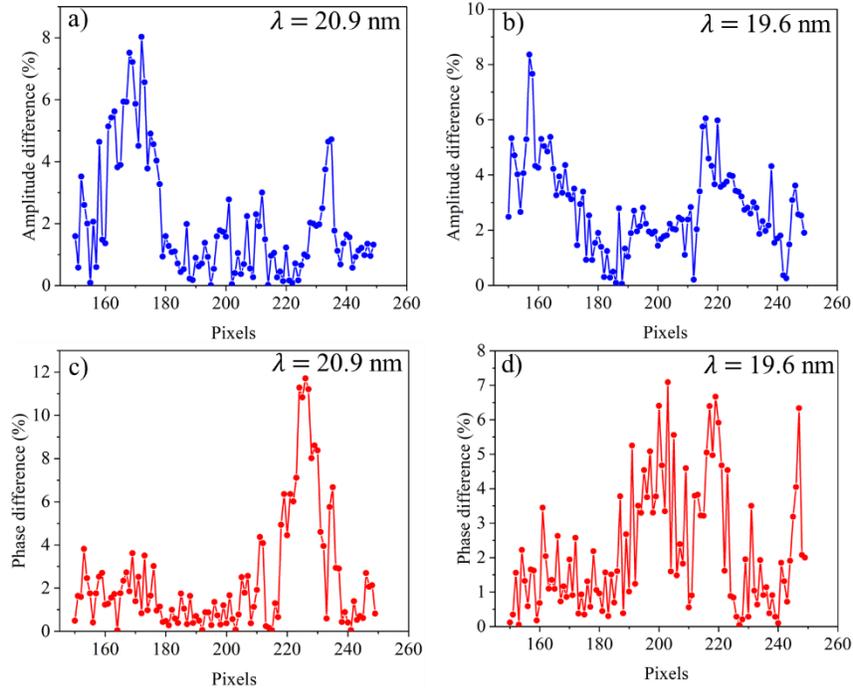

Fig. 16. Quantitative evaluation of the reconstructed spectral amplitudes and phases of the sample obtained from the simulation shown in Fig. 14. (a), (b) the difference of reconstructed spectral amplitudes and (c), (d) corresponding difference of reconstructed spectral phase from the actual values. The data is taken across the yellow dashed line shown in Fig. 15a.

Experiments of ptychography are also performed using probes focused by mirrors or zone plates. Therefore, another simulation was performed to investigate whether the method used in this study can be applied under the illumination of experimentally focused probes. The simulation was performed using the complex sample shown in Fig. 3. Meanwhile, an experimental probe focused by a Fresnel zone plate [41] was used for the illuminating probe. As the purpose of this simulation is to examine the reliability of this method for a type of focused probes in realistic conditions, the differences in the experimental and physical properties between the experiments to obtain this probe and the current study are neglected. Figures 17a and 17b respectively show the amplitude and phase of the probe for the central component in the broad spectrum. The spectral probes were generated using the scaling method described in Sec. 2.3. Figure 17c shows the broadband diffraction intensity from this type of probe at a scanning position obtained using the method described in Sec. 2.4.

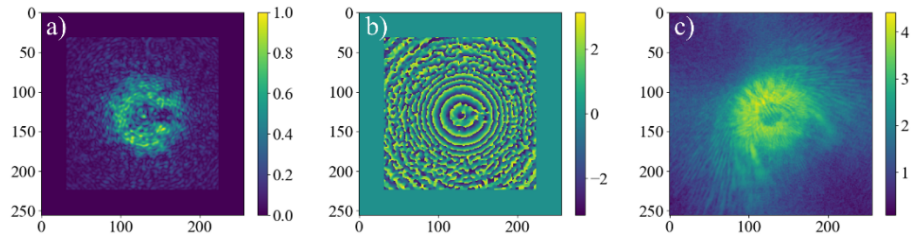

Fig. 17. Simulation with an experimentally focused probe. (a) and (b) amplitude and phase of the probe at the central wavelength, (c) broadband diffraction intensity at a scanning position using the probe shown in (a) and (b) with the complex sample shown in Fig. 3. The diffraction intensity in (c) is shown in logarithmic scale.

Figures 18a and 18b show the reconstructed amplitude and phase of the sample, respectively, for the central component wavelength of 20.3 nm. Figures 18c and 18d indicate their quantitative differences from the corresponding actual values. Other corresponding results are shown in Fig. S23 of the supplemental document. These results demonstrate the reliability of the reconstructed images using an experimentally focused probe. In fact, the structure of sample is well recovered as shown in Figs. 18a and 18b, and the spectral information of sample agrees well with the actual values, shown in Figs. 18c and 18d. The recovered spectral bandwidth coincides with the actual bandwidth, shown in Fig. S24 of the supplemental document. Therefore, it can be concluded that the reconstructed results are not significantly affected by the experimentally focused probe and the method used in this study is reliably applicable in realistic experiments with focused beams.

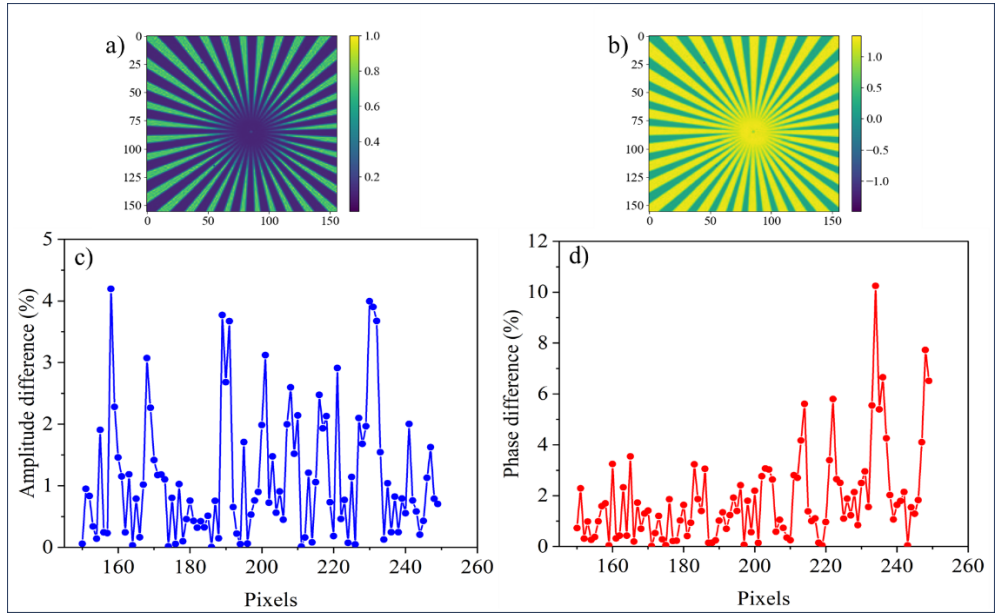

Fig. 18. Simulation with an experimental focused probe. (a) and (b) reconstructed amplitude and phase of the sample for the central wavelength of 20.3 nm, (c) and (d) quantitative differences of the reconstructed amplitude (a) and phase (b) from the actual values. The data is taken from the line profile shown in Fig. 6c.

## 4. Conclusion

This study demonstrates that high-quality images of samples can be obtained using broadband sources such as HHG sources. The detrimental effects of the broad bandwidth in diffraction imaging can be tolerated using the ptychographic approach. The obtained results also show that complex samples with amplitudes and phases can be spectrally reconstructed from broadband diffraction intensities. This presents many opportunities to apply broadband attosecond pulses to the imaging of complex samples. Furthermore, time-resolved imaging of complex samples using broadband attosecond pulses is an interesting subject for future research, by combining the method used in this study with the recently developed single-shot ptychography [42,43]. Such a technique may facilitate the imaging of ultrafast phenomena such as spin, charge, acoustics, and heat transport [44] with high-temporal and spatial resolution.

The results obtained in this research also demonstrate that the high energy IAP [15,16] can be applied in ultrafast imaging, such as ultrafast ptychography or single-shot ptychography to image complex samples, paving the new way for the retrieval of high-quality images with significant improvement in temporal resolution.

Spectral bandwidth information can also be reliably recovered along with the images of the sample. The information of bandwidth is necessary due to the dependency of diffraction intensities on the bandwidth in CDI. Therefore, this is meaningful when performing experiments with fluctuating or unknown spectral models of sources that require other experiments for beam characterization.

In addition, spectral information recovered by a single broadband ptychographic experiment can be combined with data acquired from other methods, such as XAFS, to characterize the physical and chemical properties of samples, instead of repeating experiments at different energies. Consequently, the radiation dose on samples can be significantly reduced for dose-sensitive samples, especially biological samples. Moreover, the stability of the experimental system is better with a shorter accumulation time.

**Funding.** Ministry of Education, Culture, Sports, Science and Technology of Japan (MEXT) through Grants-in-Aid under Grant No. 21H01850; and the MEXT Quantum Leap Flagship Program (Q-LEAP) (Grant No. JP-MXS0118068681).

**Acknowledgments.** We acknowledge Center for Computational Materials Science, Institute for Materials Research, Tohoku University for the use of MASAMUNE-IMR (Project No. 202212-SCKXX-0011). The authors thank Dr. Bing Xue for valuable comments and discussion.

**Disclosures.** The authors declare no conflicts of interest.

**Data availability.** Data underlying the results presented in this paper are not publicly available at this time but may be obtained from the authors upon reasonable request.

**Supplemental document.** See Supplement 1 for supporting content.

**References**

1. J. Miao, T. Ishikawa, I. K. Robinson, *et al.*, "Beyond crystallography: Diffractive imaging using coherent x-ray light sources," Science **348**(6234), 530-535 (2015).
2. I. K. Robinson, I. A. Vartanyants, G. J. Williams, *et al.*, "Reconstruction of the shapes of gold nanocrystals using coherent x-ray diffraction," Phys. Rev. Lett. **87**(19), 195505/1–195505/4 (2001).
3. J. Nelson, X. Huang, J. Steinbrener, *et al.*, "High-resolution x-ray diffraction microscopy of specifically labeled yeast cells," Proc. Natl Acad. Sci. USA **107**(16), 7235–7239 (2010).
4. J. R. Fienup, "Phase retrieval algorithms: a comparison," Appl. Opt. **21**(15), 2758–2769 (1982).
5. A. Rundquist, C. G. Durfee III, Z. Chang, *et al.*, "Phase-matched generation of coherent soft x-rays," Science **280**(5368), 1412-1415 (1998).
6. R. A. Bartels, A. Paul, H. Green, *et al.*, "Generation of spatially coherent light at extreme ultraviolet wavelengths," Science **297**(5580), 376-378 (2002).
7. E. J. Takahashi, T. Kanai, K. L. Ishikawa, *et al.*, "Coherent water window x ray by phase-matched high-order harmonic generation in neutral media," Phys. Rev. Lett. **101**(25), 253901/1-253901/4 (2008).
8. R. L. Sandberg, A. Paul, D. A. Raymondson, *et al.*, "Lensless diffractive imaging using tabletop coherent high-harmonic soft x-ray beams," Phys. Rev. Lett. **99**(9), 098103/1-098103-4 (2007).
9. A. Ravasio, D. Gauthier, F. R. N. C. Maia, *et al.*, "Single-shot diffractive imaging with a table-top femtosecond soft x-ray laser-harmonics source," Phys. Rev. Lett. **103**(2), 028104/1-028104/5 (2009).
10. K. B. Dinh, H. V. Le, P. Hannaford, *et al.*, "Coherent diffractive imaging microscope with a high-order harmonic source," Appl. Opt. **54**(17), 5303-5308 (2015).
11. P. M. Paul, E. S. Toma, P. Breger, *et al.*, "Observation of a train of attosecond pulses from high harmonic generation," Science **292**(5522), 1689-1692 (2001).
12. E. J. Takahashi, P. Lan, O. D. Mücke, *et al.*, "Attosecond nonlinear optics using gigawatt-scale isolated attosecond pulses," Nat. Commun. **4**(2691), 1-9 (2013).
13. S. M. Teichmann, F. Silva, S. L. Cousin, *et al.*, "0.5-keV soft x-ray attosecond continua," Nat. Commun. **7**(11493), 1-6 (2016).
14. J. Li, X. Ren, Y. Yin, *et al.*, "53-attosecond x-ray pulses reach the carbon K-edge," Nat. Commun. **8**(186), 1-5 (2017).
15. B. Xue, Y. Tamaru, Y. Fu, *et al.*, "Fully stabilized multi-TW optical waveform synthesizer: Toward gigawatt isolated attosecond pulses," Sci. Adv. **6**(16), 1-10 (2020).
16. B. Xue, K. Midorikawa, E. J. Takahashi, "Gigawatt-class, tabletop, isolated-attosecond-pulse light source," Optica **9**(4), 360-363 (2022).
17. K. Midorikawa, "Progress on table-top isolated attosecond light sources," Nat. Photon. **16**, 267-278 (2022).
18. H. Wen, M. J. Cherukara and M. V. Holt, "Time-resolved X-ray microscopy for materials science," Annu. Rev. Mater. Res. **49**(1), 389-415 (2019).

19. T. Helk, M. Zürch, C. Spielmann, "Perspective: Towards single shot time-resolved microscopy using short wavelength table-top light sources," Struct. Dyn. **6**(1), 010902/1-010902/10 (2019).
20. B. Abbey, L. W. Whitehead, H. M. Quiney, *et al.*, "Lensless imaging using broadband X-ray sources," Nat. Photon. **5**, 420-424 (2011).
21. E. Malm, H. Wikmark, B. Pfau, *et al.*, "Singleshot polychromatic coherent diffractive imaging with a high-order harmonic source," Opt. Express **28**(1), 394-404 (2020).
22. J. Huijts, S. Fernandez, D. Gauthier, *et al.*, "Broadband coherent diffractive imaging," Nat. Photon. **14**, 618-622 (2020).
23. W. Eschen, S. Wang, C. Liu, *et al.*, "Towards attosecond imaging at the nanoscale using broadband holography-assisted coherent imaging in the extreme ultraviolet," Commun Phys **4**(154), 1-7 (2021).
24. J. M. Rodenburg, "Ptychography and related diffractive imaging methods," Adv. Imaging Electron Phys. **150**, 87-184 (2008).
25. F. Pfeiffer, "X-ray ptychography," Nat. Photon. **12**, 9-17 (2018).
26. D. J. Batey, D. Claus, J. M. Rodenburg, "Information multiplexing in ptychography," Ultramicroscopy **138**, 13-21 (2014).
27. A. Rana, J. Zhang, M. Pham, *et al.*, "Potential of attosecond coherent diffractive imaging," Phys. Rev. Lett. **125**(8), 086101/1-086101/6 (2020).
28. Y. Yao, Y. Jiang, J. Klug, *et al.*, "Broadband x-ray ptychography using multi-wavelength algorithm," J. Synchrotron Rad. **28**, 309-317 (2021).
29. R. Liu, W. Cao, Q. You, *et al.*, "Broadband ptychographic imaging with an accurately sampled spectrum," Phys. Rev. A **107**(3), 033510/1-033510/9 (2023).
30. B. Abbey, G. J. Williams, M. A. Pfeifer, *et al.*, "Quantitative coherent diffractive imaging of an integrated circuit at a spatial resolution of 20 nm," Appl. Phys. Lett. **93**(21), 214101/1-214101/3 (2008).
31. B. L. Henke, E. M. Gullikson, J. C. Davis, "X-ray interactions: photoabsorption, scattering, transmission, and reflection at E = 50-30,000 eV, Z = 1-92", At. Data Nucl. Data Tables **54**(2), 181-342 (1993).
32. M. Born and E. Wolf, "Principles of Optics," Cambridge University Press (1999).
33. A. M. Maiden, J. M. Rodenburg, "An improved ptychographical phase retrieval algorithm for diffractive imaging," Ultramicroscopy **109**(10), 1256-1262 (2009).
34. M. Pham, A. Rana, J. Miao, *et al.*, "Semi-implicit relaxed Douglas-Rachford algorithm (sDR) for ptychography," Opt. Express **27**(22), 31246-31260 (2019).
35. J. N. Clark, X. Huang, R. J. Harder, *et al.*, "Dynamic imaging using ptychography," Phys. Rev. Lett. **112**(11), 113901/1-113901/5 (2014).
36. F. van der Veen, F. Pfeiffer, "Coherent x-ray scattering," J. Phys.: Condens. Matter **16**(28), 5003-5030 (2004).
37. C. Jacobsen, J. Deng, Y. Nashed, "Strategies for high-throughput focused-beam ptychography," J. Synchrotron Rad. **24**(5), 1078-1081 (2017).
38. H. Uematsu, N. Ishiguro, M. Abe, *et al.*, "Visualization of structural heterogeneities in particles of lithium nickel manganese oxide cathode materials by ptychographic X-ray absorption fine structure," J. Phys. Chem. Lett. **12**(24), 5781-5788 (2021).
39. M. Abe, F. Kaneko, N. Ishiguro, *et al.*, "Visualization of sulfur chemical state of cathode active materials for lithium-sulfur batteries by tender X-ray spectroscopic ptychography," J. Phys. Chem. C **126**(33), 14047-14057 (2022).
40. M. R. Howells, T. Beetz, H. N. Chapman, *et al.*, "An assessment of the resolution limitation due to radiation-damage in X-ray diffraction microscopy," J. Electron Spectros Relat. Phenom. **170**(1-3), 4-12 (2009).
41. K. Wakonig, H. Stadler, M. Odstrčil, *et al.*, "PtychoShelves, a versatile high-level framework for high-performance analysis of ptychographic data," J. Appl. Cryst. **53**, 574-586 (2020).
42. P. Sidorenko and O. Cohen, "Single-shot ptychography," Optica **3**(1), 9-14 (2016).
43. K. Kharitonov, M. Mehrjoo, M. R. Lopez, *et al.*, "Single-shot ptychography at a soft X-ray free-electron laser," Sci Rep **12**(14430), 1-9 (2022).
44. X. Shi, C. T. Liao, Z. Tao, *et al.*, "Attosecond light science and its application for probing quantum materials," J. Phys. B: At. Mol. Opt. Phys. **53**(18), 1-20 (2020).